# Fifty Years of Breakthrough Discoveries in Fluid Criticality

Mikhail A. Anisimov

*Department of Chemical and Biomolecular Engineering and Institute for Physical Science and Technology, University of Maryland, University of Maryland, College Park, MD 20742, USA*
(August 11, 2011)

E-mail anisimov@umd.edu   Phone (301) 405 8049   Fax (301) 314 9404

**Abstract**

Fifty years ago two scientists, who celebrate their $80^{th}$ birthdays in 2011, Alexander V. Voronel and Johannes V. Sengers performed breakthrough experiments that challenged the commonly accepted views on critical phenomena in fluids. Voronel discovered that the isochoric heat capacity of argon becomes infinite at the vapor-liquid critical point. Almost simultaneously, Sengers observed a similar anomaly for the thermal conductivity of near-critical carbon dioxide. The existence of these singularities was later proved to be universal for all fluids. These experiments had a profound effect on the development of the modern (scaling) theory of phase transitions, which is based on the diverging fluctuations of the order parameter. In particular, the discovery of the heat-capacity divergence at the critical point was a keystone for the formulation of static scaling theory, while the discovery of the divergence of the thermal conductivity played an important role in the formulation of dynamic scaling and mode-coupling theory. Moreover, owing to the discoveries made by Voronel and Sengers 50 years ago, critical phenomena in fluids have become an integral part of contemporary condensed-mater physics.

**Keywords**

Voronel, Sengers, Critical Phenomena, Fluids, Heat capacity, Thermal Conductivity

Fifty years ago, two young physicists, Alexander (Sasha) V. Voronel in the USSR National Bureau of Standards, Moscow Region and Johannes (Jan) V. Sengers in the van der Waals laboratory, Amsterdam performed breakthrough experiments that challenged the commonly accepted views on critical phenomena in fluids. The photos, Figs. 1 and 2, were shot at that time. In 2011, both scientists celebrate their 80$^{th}$ birthdays. Currently, Voronel is Professor Emeritus at the University of Tel-Aviv and Sengers is Distinguished Professor Emeritus at the University of Maryland, College Park.

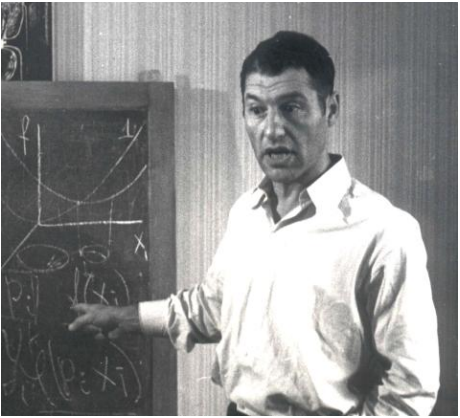

Fig.1. Alexander Voronel at a seminar talk, USSR Bureau of Standards, early 1960's.

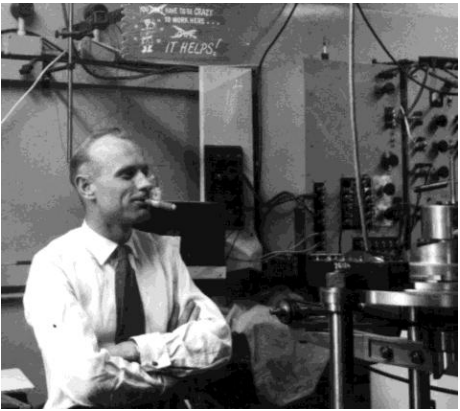

Fig.2. Jan Sengers in the van der Waals laboratory, University of Amsterdam, early 1960's.

In 1961, using an improved adiabatic calorimeter [1], Voronel and his students, Bagatskiĭ and Gusak, measured the isochoric heat capacity of argon near the vapor-liquid critical point. After a few months of carefully checking for all possible artifacts, the authors sent a short letter to JETP [2]. The results, shown in Fig. 3, shocked the Russian physics community. The authors emphasized striking similarity of the observed anomaly in the heat capacity with the singularity at the superfluid transition in liquid helium, previously reported by Buckingham and Fairbank [3]. They claimed that the isochoric heat capacity of argon diverges (becomes infinite) at the critical point along the transition from the two-phase state to the supercritical homogeneous fluid. This claim was in the striking contradiction with the van der Waals-Landau theory of critical phenomena [4], which was commonly regarded as "untouchable" at that time. The theory predicted a discontinuity of the finite isochoric heat capacity upon crossing the two-phase boundary. Commonly used in engineering practice analytical equations of state, from the van der Waals equation to most sophisticated ones, are based on this classical theory and all predict the finite isochoric heat capacity at the critical point.

A footnote in the 1964 Russian edition of Landau and Lifshitz "Statistical Physics" [5], which correctly stated that the isochoric heat-capacity divergence has no grounds in the classical, mean-field, theory of phase transitions, was wrongly interpreted by many as a fatal criticism of Voronel's conclusion [6]. However, Voronel's discovery was immediately recognized by a bright young theorist at King's College in London, Michael E. Fisher [7], who believed in the universal nature of the apparently very different phase transitions, such that in liquid helium, binary alloys, ferromagnetic and ferroelectric materials, and in classical fluids. Professor Fisher, currently at the University of Maryland, College Park, also celebrates his 80[th] birthday in 2011.

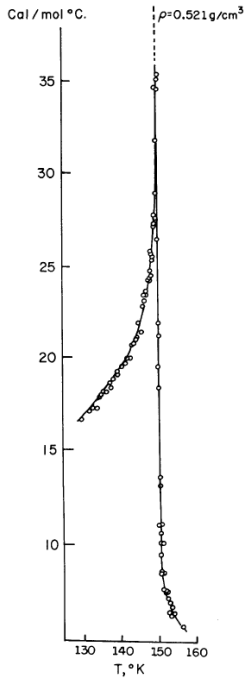

Fig.3. Variation of the isochoric heat capacity of argon along the critical isochore [2].

In the Richard T. Cox Lecture at the 2006 APS Annual Meeting, Voronel noted: "Now it is even strange to think that in the 1950s the second-order phase transitions and the liquid-vapor critical point were considered as different fields of physics" [8]. In fact, for a large part of physics community, especially in the United States and to some degree in Russia, the liquid state was not part of condensed-matter physics at all, actually belonging to engineering or, at best, to physical chemistry [9].

One notable exception was in the Netherlands, where the traditions of van der Waals and Kamerlingh Onnes continued to flourish, through the combination of the most precise experimental techniques with comprehensive analysis, in the studies of fluids [10,11]. Remarkably, at about the same time, when Voronel challenged the classical thermodynamics of vapor-liquid critical point, a doctoral researcher at the van der Waals laboratory, Jan Sengers, confronted the established classical views on transport properties of fluids in the critical region [12]. He used a parallel plate method to measure the thermal conductivity of carbon dioxide near the vapor-liquid critical point. This method permits very small plate distances and small temperature gradients and thus is the most appropriate for near-critical states. By eliminating the effects of thermal convection and carefully studying the variations of the plate distance, the temperature gradient, and the horizontality of the layer, Sengers proved that the thermal conductivity possesses a pronounced maximum at the critical density. Moreover, the magnitude of this maximum increased upon a gradual approach to the critical isotherm suggesting that the thermal conductivity could tend to infinity at the critical point.

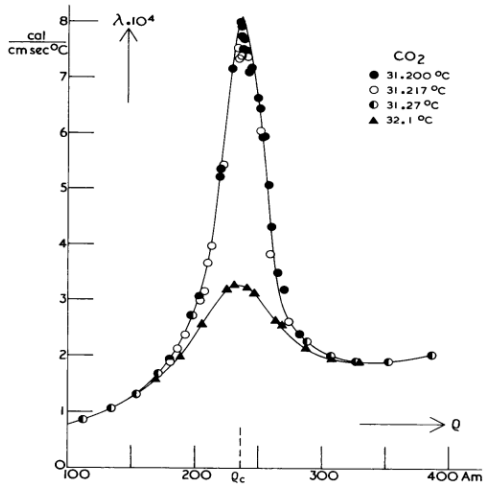

Fig.4. Density variation of the thermal conductivity of carbon dioxide along isotherms in the critical region [12]. The density is shown in Amagat units; 1 Am = 44.614774 mol/m$^3$.

It was not surprising that the results reported by Voronel and Sengers were considered by many even experienced scientists as controversial and doubtful. Criticality is a very delicate state of matter. Critical phenomena in fluids fascinate everyone who encounters them – one only needs to think of a system in which mutual diffusion of species practically stops, a sound wave is damped after traveling a distance of only a few wavelengths, a laser beam is diffused during the path through an optical cell, heat capacity and thermal conductivity diverge, and thermal perturbation do not relax for many hours or even days. However, reliable experiments near critical points of fluids are extremely challenging to conduct. Experimentalists always have to fit the results of their measurements to an idealized theoretical model. The specific and unavoidable feature of critical phenomena is that, owing to the huge susceptibility, even small perturbations associated with the measurements may lead to a dramatic distortion, thus result in a qualitative change in the observed anomalies [13,14]. Thus the quality and reproducibility of experimental data collected in the critical region are often determined by the physical state of the system under investigation rather than by the resolution of the instrument. Reproducible distortions of "ideal" critical anomalies have often been sources of misinterpretation and confusion. In a typical thermal experiment on a fluid, the perturbation may be neglected if

$$\frac{\varepsilon}{k_\mathrm{B} T} \ll 1, \quad (1)$$

where $\varepsilon$ is the perturbation energy per molecule and $k_\mathrm{B}$ is Boltzmann's constant. In the critical region, where the susceptibility of fluids to external perturbations is anomalously large, the condition (1) is insufficient and must be replaced by a much more severe constrain:

$$\frac{\varepsilon}{k_\mathrm{B}T} \ll |\tau|, \qquad (2)$$

where $\tau = (T - T_\mathrm{c})/T_\mathrm{c}$ with $T_\mathrm{c}$ being the critical temperature. At a distance from the critical temperature (along the critical isochor) of about 0.3 K with $T_\mathrm{c} \cong 300$ K, the requirement for obtaining undisturbed data is a thousand times more severe than at regular conditions! This is similar to the difficulty of performing accurate experiments at extremely low temperatures.

Another important constraint is ensuring that the near-critical fluid is in thermodynamic equilibrium (for the measurements of thermodynamic properties) or in a well controlled steady state (for the measurements of transport properties). Relaxation to equilibrium or to the steady state is slow in the critical region. The thermal relaxation time in a single-component near-critical fluid can be estimated as

$$t_\mathrm{TR} \cong \frac{l^2 \rho C_P}{\lambda}, \qquad (3)$$

where $l$ a characteristic length scale (*e. g.* the linear size of the calorimeter cell), $\rho$ is the density, and $\lambda$ is the thermal conductivity. Since, upon approaching the critical point, the isobaric heat capacity increases much faster than the thermal conductivity, the relaxation time increases and ultimately diverges at the critical point. To obtain thermodynamic equilibrium data in the critical region, one must be sure that the relaxation time is much smaller than the characteristic measurement time. Other dangerous disturbing factors are gravity (especially for heat-capacity measurements) and thermal convection (especially for the thermal-conductivity measurements). Due to the anomalously large compressibility, the presence of gravity causes a significant inhomogeneity of the density and may dramatically affect experiments in the critical region [13,15]. If in a near-critical fluid the temperature is not uniform, gravity can easily generate convection and convective heat flow. In the experimental design, the latter must be distinguished from the irreversible fluxes associated with the thermodynamic forces [12,16].

The experiments of Voronel and Sengers were unique because they were the first who systematically formulated and implemented in practice the scientific requirements for obtaining undisturbed and reliable experimental information on near-critical fluids. These experiments had a profound effect on the development of the modern (scaling) theory of phase transitions, which is based on the diverging fluctuations of the order parameter. In particular, the discovery of the heat-capacity divergence at the critical point was a keystone for the formulation of static scaling theory [17-19], while the discovery of the divergence of the thermal conductivity played a crucial role in the formulation of dynamic scaling and mode-coupling theory [20-23]. "In fact, this combination of theory and experiments ultimately made the science of fluids an accepted part of modern condensed-matter physics.

With an increase in the resolution and overall quality of experiments, it became obvious by the early 1970's that the divergence of the isochoric heat capacity and thermal conductivity at the gas-liquid critical point are universal phenomena for all single-component fluids [13,16]. All fluids and fluid mixtures belong to the Ising-model class of universality in statics and to the conserved-order-parameter universality class in dynamics. This universality is associated with the universal nature of critical fluctuations [24,25]. What distinguishes the isochoric heat capacity and thermal conductivity among other thermodynamic and transport properties of fluids is that their singularities, discovered by Voronel and Sengers, are completely determined by the divergence of the density fluctuations at the critical point. Classical (mean-field) theory neglects these fluctuations and predicts a finite heat capacity and a finite thermal conductivity. Specific power laws that describe the anomalies of the isobaric heat capacity, isothermal compressibility, or volumetric expansivity are affected by fluctuations; however their divergence is required by general thermodynamics, even in the absence of fluctuations. Similarly, the thermal diffusivity, $D = \lambda / \rho C_P$, vanishes at the critical point with or without accounting for fluctuations: the diverging thermal conductivity only changes the power law that control the vanishing diffusivity.

According to the modern theory of critical phenomena, supported by most accurate light-scattering experiments, the "size" of fluctuation inhomogeneities, $\xi$, known as the correlation length, diverges at the critical point. Along the critical isochore, asymptotically close to the critical point,

$$\xi \propto |\tau|^{-\nu}, \tag{4}$$

where $\nu = 0.630$ is a critical exponent, calculated by renormalization-group theory for the Ising-model universality class [25,26]. Scaling theory predicts that the singularity of the isochoric heat capacity is associated with the divergence of the correlation length as

$$C_V \propto \frac{\partial^2 \xi^{-3}}{\partial \tau^2} \propto |\tau|^{-(2-3\nu)}. \tag{5}$$

Thus the theory predicts a power law for the divergence of the heat capacity with a critical exponent $\alpha = 2 - 3\nu = 0.11$. This prediction is fully supported by the most accurate experiments [26,27].

Similarly, dynamic scaling and mode coupling theory [20-23] predict that the thermal diffusivity near the critical point should obey the Stokes-Einstein equation:

$$D = \frac{\lambda}{\rho C_P} = \frac{k_B T}{6\pi \mu \xi}, \tag{5}$$

where $\mu$ is the shear viscosity. Since, according to static scaling, the isobaric heat capacity strongly diverges along the critical isochore as

$$C_P \propto \xi^{2-\eta} \propto |\tau|^{-(2-\eta)\nu}, \tag{6}$$

with a critical exponent $\eta \cong 0.03$ [26]. Hence, the thermal conductivity diverges as

$$\lambda \propto \frac{\xi^{1-\eta}}{\mu}. \tag{7}$$

Sengers noted in Summary of his Ph.D. Thesis [12]: "The similarity in the behaviour of $\lambda$ and $C_V$ suggests that the sharp increase of $\lambda$ and $C_V$ in the critical region is connected with the same fundamental process".

Even after Eqs. (5) and (7) were theoretically established, there was a misinterpretation of the asymptotic behavior of the diffusivity and thermal conductivity because the analytic background of the thermal conductivity was neglected. Sengers and his student Keyes resolved the apparent contradiction between theory and experiment and proved the validity of Eq. (5) [28]. More recently, it was shown that the shear viscosity also exhibits a singularity, though extremely week [29], that should be included in the asymptotic analysis of transport properties in the critical region.

The concept of critical-point universality was extended to fluid mixtures via the principle of isomorphism. Voronel was a co-author of an earlier formulation of isomorphism [30], followed later by a more advanced formulation [31] and one that also included transport properties [32].

Owing to the discoveries made by Voronel and Sengers 50 years ago, critical phenomena in fluids and fluids mixtures have become an integral part of condensed-mater physics. These fascinating phenomena are universally described by the elegant theory of mesoscopic fluctuations and strongly supported by beautiful experiments.

**Acknowledgements**

I acknowledge a long-term friendship and research collaboration (1968-1974) with Sasha and with Jan (since 1984). I also thank Nina Voronel and Victor Steinberg for sending me photos from the 1960's.